\documentclass{article}
\usepackage{spconf,amsmath,graphicx}

\title{Unsupervised heart abnormality detection based on phonocardiogram analysis with Beta Variational Auto-Encoders}
\name{Shengchen Li, Ke Tian, Rui Wang}
\address{Beijing University of Posts and Telecommunications}
\begin{document}
%\ninept
%
\maketitle
\begin{abstract}
Heart Sound (also known as phonocardiogram (PCG)) analysis is a popular way that detects cardiovascular diseases (CVDs). Most PCG analysis uses supervised way, which demands both normal and abnormal samples. This paper proposes a method of unsupervised PCG analysis that uses beta variational auto-encoder ($\beta-\text{VAE}$) to model the normal PCG signals. The best performed model reaches an AUC (Area Under Curve) value of 0.91 in ROC (Receiver Operating Characteristic) test for PCG signals collected from the same source. Unlike majority of $\beta-\text{VAE}$s that are used as generative models, the best-performed $\beta-\text{VAE}$ has a $\beta$ value smaller than 1. Further experiments then find that the introduction of a light weighted KL divergence between distribution of latent space and normal distribution improves the performance of anomaly PCG detection based on anomaly scores resulted by reconstruction loss. The fact suggests that anomaly score based on reconstruction loss may be better than anomaly scores based on latent vectors of samples.
\end{abstract}
\begin{keywords}
Phonocardiogram Analysis, Variational-Auto-Encoder, Anomaly Detection, Outlier Detection, Unsupervised Learning
\end{keywords}
\section{Introduction}
\label{sec:intro}

With the rapid development of smart wearing and home care devices, the phonocardiogram (PCG) analysis is used to identify the risk of cardiovascular diseases (CVDs) due to more portable devices requirements and easier collection of samples. Most existing methods require a collection of both normal and abnormal samples to build a supervised machine learning system. Collecting abnormal samples could be a tricky task as the distribution of CVD types leads to a possible bias of the data, which potentially harms the system performance. Moreover the labelling attempts of PCG signals are professionally labour expensive. This paper proposes an unsupervised system with normal PCG signals only, which detects potential risks of CVDs with less labelling attempts and easier dataset design.

The proposed system attempts to analyse normal PCG signals with the most distinguishable features extracted. As there are no abnormal samples used in the training process of the proposed system, the main task of the proposed system is to find a type of representation for PGC signals such that normal signals can be represented by similar feature vectors. As a result, Variational Auto-Encoders (VAE) is proposed to encode the normal PCG signals where the latent vectors in the latent space of VAE follow Gaussian distribution. 

$\beta-$VAEs are used for generative models in most cases, which balance the cost of reconstruction accuracy and distribution in latent space~\cite{burgess_understanding_2018}. As a result, $\beta-\text{VAE}$ is a variation of VAE that introduces a variable $\beta$ to find the right balance between the distribution of latent space and reconstruction loss. Serving as a generative model, $\beta-$ VAEs usually introduce a value of $\beta$ larger than 1. But in this paper, $\beta-$VAEs are used for outlier detector, it is still worthy to discuss the value of $\beta$ for the best performed system. With the PhysioNet / CinC Dataset~\cite{liu_open_2016}, VAE architectures with different value of beta are tested.

As most outlier detection systems use either reconstruction loss or variables related to latent vectors to calculate anomaly score. This paper further explores how the KL divergence between the latent vectors and the reconstruction loss are related to each other. Based on the results, this paper investigates the preferred variable for anomaly score calculation in the PCG analysis.

The following sections of the paper is organised by the following way. Related works are reviewed first. Then the basic architecture of the proposed system is introduced. The experiment setup is then explained with the results followed. The results are then discussed followed by a conclusion. 

\section{Literature Review}
\label{sec:background}

Phonocardiograms, or PCGs are of great value, for which can be used to detect heart disease. Related attempts can be dated back to the year of 1995~\cite{barschdorff1995automatic}. In the year of 2016, PhysioNet~\cite{physionet} and CinC (Computing in Cardiology Challenge) organised a data challenge that detects anomaly PCG signals. Since then, the release of the challenge dataset~\cite{liu_open_2016} promotes the research in relevant topics. 

Traditional methods, such as support vector machine~\cite{zabihi_heart_2016}, i-vector~\cite{adiban2019statistical} and hidden Markov model \cite{grzegorczyk_pcg_2016} are used to detect anomaly PCG signals in a supervised way. Deep learning methods based on Variational Auto-Encoder (VAE)~\cite{banerjee2020semi}, deep convolutional neural network~\cite{koike2020audio,rubin_recognizing_2017} and recurrent neural network~\cite{qian2019deep} are also used recently. Assembling of both traditional methods and deep learning methods, the best performed system~\cite{potes_ensemble_2016} of the challenge achieves an accuracy of 0.91.

Most existing methods uses a supervised system such that collecting large scale of data is more difficult and the labelling process is professionally labour expensive. As a result, there are some attempts on unsupervised system. Unnikrishnan et al.~\cite{unnikrishnan_semi-supervised_2020} use auto-encoder to distinguish normal and abnormal PCG signals with an AUC of 0.828, which needs further improvements compared with the state-of-the-art systems.

As a result, this proposes a VAE based method. VAE system does not only pursue an accurate reconstruction of original signals but also require the distribution of latent space obeys a specific probability distribution (usually normal distribution). Existing works show that VAE may outperform the more traditional AE based system for anomaly detection~\cite{yao_unsupervised_2019}. However, as VAEs are used as generative models in most cases, this paper proposes to use $\beta$-VAE~\cite{higgins_-vae_2017} that enables controls on the balance between the reconstruction loss and the distribution in latent space. 

Moreover, most existing works consider the dataset as a single domain problem, i.e. there are no samples that collected by a different way used in the evaluation process. To extend the application of the proposed system, this paper considers the PCG anomaly detection problem as a multiple domain problem. Namely the data collected for training and the data used for testing are collected by different devices. Particularly in this paper, unlike Banerjee and Ghose~\cite{banerjee2020semi}, this paper uses different subsets in the Physio / CinC dataset, which is more challenging in terms of data inconsistency. 

\section{Methods}
\label{sec:methods}

In this section, the system architecture is introduced, which has three stages: pre-processing, VAE and post-processing. The pre-processing stage normalise Mel Spectrogram along frequency bins. The VAE system encodes and decodes the normalised Mel Spectrogram splits to produce raw anomaly score with reconstruction loss. which is used to calculate the final anomaly score of a piece of audio at post-processing stage.

\subsection{Dataset}

The Dataset used in the proposed method is the Physio Heart Sound Dataset~\cite{liu_open_2016}, in which six subsets of data are collected from different data source. The number of samples and the positive rate are listed in Table~\ref{tab:Physio}. Each subset are collected at different locations with difference sample-collect devices~\cite{liu_open_2016} hence can be considered as different data domain. 

\begin{table}[hbt]
	\centering
	\begin{tabular}{|c|c|c|c|c|c|c|}
		\hline
		Subset & a & b & c & d & e & f \\
		\hline
		Sample \# & 409 & 490 & 31 & 55 & 2141 & 114 \\
		\hline
		Positive \% & 0.71 & 0.21 & 0.78  & 0.51 & 0.09 & 0.30 \\
		\hline
	\end{tabular}
	\caption{The sample number and positive (abnormal) rate in each subset of Physio Heart Beat dataset. }
	\label{tab:Physio}
\end{table}

\subsection{Pre-Processing}

The Physio Heart Sound datasets contain samples with variable lengths which range from 5 seconds to 120 seconds. All samples are pre-processed to last 8 seconds for containing 6 to 13 complete cardiac cycles. Shorter samples are padded by a recurrent manner and Longer samples are simply truncated. 

Then the Mel Spectrogram is calculated by formulae specified by Davis and Mermelstein \cite{Davis1980} with a window size of 1024, hope size of 512 and 14 Mel filters applied. Given the sampling rate of 2 kHz, each frame last about 0.51 seconds. 

Suppose Mel Spectrogram is represented by $\mathbf{S}_{M\times N}$, where $M$ is the number of frequency bins and $N$ is the number of frames. The Mel Spectrogram is then normalised regarding to each frames. Using $s_{m, n}$ to represent the element at $m$th row and $n$th column in spectrogram   $\mathbf{S}_{M\times N}$ i.e. the normalised $m$th frequency bin $\hat{\mathbf{S}}_m$ can be calculated as 

\begin{equation}
	\hat{\mathbf{S}}_m = \frac{s_{m, n}-mean(s_{m, n})}{std(s_{m, n})}.
\end{equation}

The resulting normalised Mel Spectrogram is then divided into super-frames. Each super-frame is formed by five consecutive frames hence last 3.07 seconds, which should be able to contain a few heart beats in most cases . The start frame of super-frames has a hop size of a single frame. Namely, for a piece of audio that has $N$ frames, there are $N-4$ super-frames. 

\subsection{VAE Architecture}

The input of the VAE system is the super-frames of the normalised Mel Spectrogram. In the training process of the proposed system, each batch contains 640 super-frames (i.e. five consecutive frames) with batch normalisation applied. 

\begin{figure*}[t]
	\includegraphics[width=\textwidth]{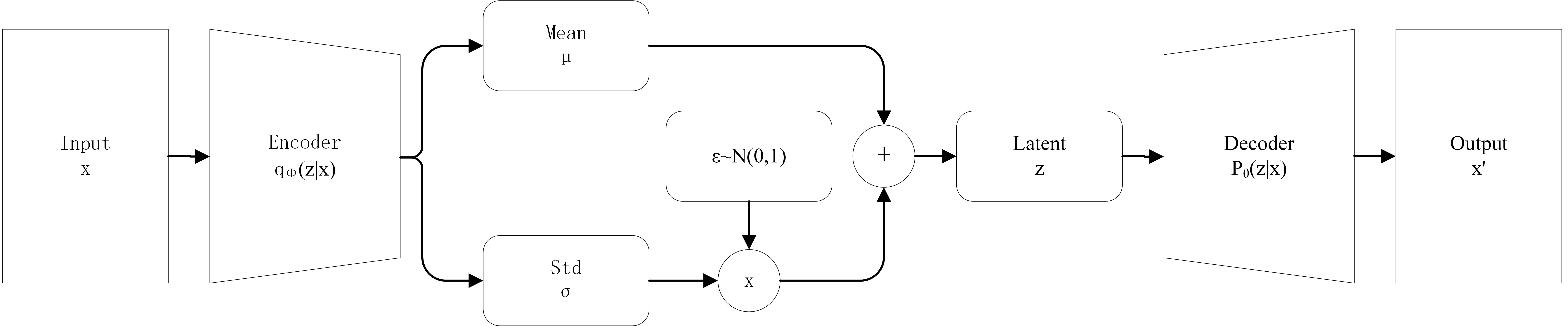}
	\caption{VAE architecture overview}
	\label{fig:vae}
\end{figure*}

The encoder of the proposed VAE system has a simple structure of four fully-connected hidden layers who has 32, 32, 16 and 16 neurons respectively. The resulting latent representations have 16 dimensions, which are regarded as the output layer of the encoder and the input layer of the decoder. The decoder has the exact structure but with reverse order in terms of number of neurons in each layer and reversed input and output layer. Figure \ref{fig:vae} shows the proposed VAE architecture. 

Following the loss function of $\beta-$VAE, the loss function of the whole VAE is the sum of reconstruction loss and the KL divergence between the resulting latent variable distribution and the normal distribution. Particularly in the proposed system, the reconstruction loss is proposed to be used as the anomaly scoring variable hence entropy-based measurements such as Evidence Lower Bound (ELBO) cannot be used as the reconstruction loss. Instead the proposed system use Mean Squared Error (MSE) as reconstruction loss function. 

Suppose $\overline{\hat{S}_{M\times N}}$ represents the mean value of normalised Mel Spectrogram, there are $K$ dimensions in the latent space. The loss function $\mathcal{L}$ used in the proposed system can be represented as 

\begin{equation}
\begin{aligned}
		\mathcal{L} = \frac{1}{NM}\sum_{m=1}^{M}\sum_{n=1}^{N}(s_{m,n}-\overline{\hat{S}_{M\times N}})^2 \\ - 0.5\beta \sum_{k=1}^{K}(1+\ln{\sigma_k^2}-\mu_k^2-\sigma_k^2).
\end{aligned}
\end{equation}

\subsection{Post-Processing}

The anomaly score of each super-frame in a piece of audio is then averaged to be used as the final anomaly score for the audio. As the super-frames have a length of 5 raw frames, there are $N-4$ super-frames in a piece of audio with $N$ frames. Using $a_i$ to represent the anomaly score of the $i$th super-frame, the overall anomaly score $a$ is then can calculated as

\begin{equation}
	a = \frac{1}{N-4}\sum_{i=1}^{N-4}a_i.
\end{equation}

\section{Results}
\label{sec:results}

\subsection{Single Subset System}

Firstly, the system performance with a single subset of data is investigated. As there are very few normal samples, the subset `C' is not used. For other subsets (subset `a', `b', `d', `e', `f'), 90\% of normal data is used to train the $\beta-$VAE model. The remaining normal data and all abnormal data is used to evaluate the performance of resulting system. There are six proposed settings for the value of $\beta$: 0, 0.01, 0.1, 1, 10, 100. When $\beta=1$, the system is essentially a common VAE system but when $\beta=0$, the system is not an auto-encoder system as an extra sampling process is introduced in the latent space.

\begin{figure}
	\centering
	\includegraphics[width=\columnwidth]{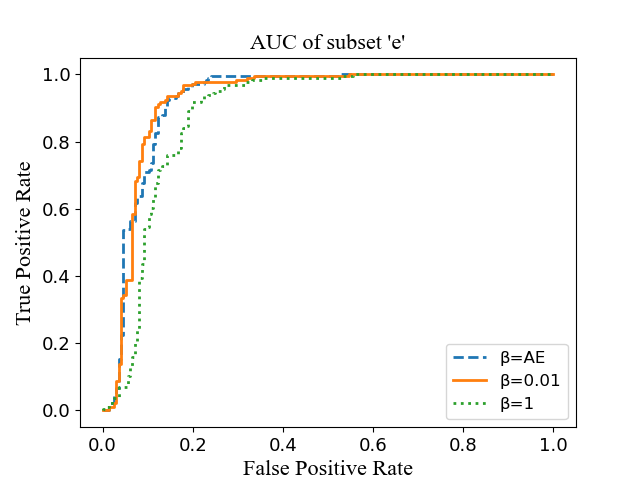}
	\caption{The Receiver Operating Characteristic (ROC) analysis that compares the true and false positive rate. The result is obtained by a system trained and tested with subset `e' in the Physio / CinC Heart Sound dataset.}
	\label{fig:roc}
\end{figure}

To evaluate the proposed systems, Receiver Operating Characteristic (ROC) analysis is performed. Area Under Curve (AUC) is calculated as the area under a curve that formed by true and false positive rate as demonstrated in Figure \ref{fig:roc}, where the result of subset `e' is used as an example. From the diagram, the VAE with $\beta=0.01$ is slightly better than the auto-encoder and both systems are better than classical VAE system (where $\beta=1$) according to the value of AUC.

In Table \ref{tab:single}, the proposed system is tested with six different settings of beta values besides a reference auto-encoder. In most cases (subsets `a', `d', `e'), $\beta-\text{VAEs}$ with a small value of $\beta$ such as 0, 0.01, 0.1 outperform other systems. For subset `b', the auto-encoder outperforms all VAEs with a preference of smaller $\beta$ value in $\beta-\text{VAEs}$. For subset `f' the system prefers a larger value of $\beta$. As a result, the overall performance of candidate systems shows that the $\beta-\text{VAEs}$ with a smaller $\beta$ value may perform 

\begin{table}[hbt]
	\centering
	\begin{tabular}{|c|c|c|c|c|c|}
		\hline
		$\beta$ & a & b & d & e & f \\
		\hline
		AE & 0.816 & 0.583  & 0.667 & 0.922 & 0.835 \\
		\hline
		0 & 0.823 & 0.579  & 0.750 & 0.924 & 0.827 \\ 
		\hline
		0.01 & 0.825 & 0.561  & 0.607 & 0.923 & 0.823 \\ 
		\hline 
		0.1 & 0.821 & 0.551  & 0.583 & 0.918 & 0.824 \\
		\hline
		1 & 0.798 & 0.559  & 0.607 & 0.881 & 0.824 \\
		\hline
		10 & 0.800 & 0.557  & 0.631 & 0.881 & 0.805 \\
		\hline 
		100 & 0.803 & 0.551  & 0.691 & 0.881 & 0.846 \\
		\hline
	\end{tabular}
	\caption{The AUC values with different settings of beta value where a single subset in Physio / CinC dataset is used for training. Larger values indicate better performance. Cross validation is applied with the same-domain test. }
	\label{tab:single}
\end{table}

Usually a larger beta value in $\beta-$VAE gives a higher priority to the distribution of latent space with a cost of reconstruction loss. In the extreme case that an auto-encoder is used, the system does not consider the distribution of latent space at all. This experiment demonstrates that the introduction of loss for latent space distribution helps the anomaly detection with unmet data. Such improvements only cost on little accuracy of anomaly detection with the in-domain data.

\subsection{Multiple Subsets System}

Next the proposed experiments is repeated with multiple subsets. The multiple domain tests introduce a high complexity of data distribution. We perform the experiments with three sets of subset combination: `ae', `ef' and all subsets (`abcdef'). Similar with the case of single domain system, in all cases, 90\% of normal data is used for training and the remaining 10\% of normal data and all abnormal data are used for testing.

\begin{table}
	\centering
	\begin{tabular}{|c|c|c|c|}
		\hline
		$\beta$ & `ae' & `ef' & all \\
		\hline
		AE & 0.820 & 0.847 & 0.783 \\
		\hline
		0 & 0.823 & 0.890 & 0.782 \\
		\hline
		0.01 & 0.822 & 0.899 & 0.786 \\
		\hline
		0.10 & 0.810 & 0.891 & 0.750 \\
		\hline
		1 & 0.766 & 0.838 & 0.678 \\
		\hline
		10 & 0.765 & 0.836 & 0.678 \\
		\hline 
		100 & 0.765 & 0.836 & 0.644 \\
		\hline 
	\end{tabular}
	\caption{The AUC values with different settings of beta value where the three sets of subset combinations in Physio / CinC dataset used for training. Larger values indicate better performance. Cross validation is applied with the same-domain test. 
	\label{tab:multiple}}
\end{table}

In Table \ref{tab:multiple}, the best performed system for subsets `ae' has a $\beta$ value of 0 and the best performed system has a $\beta$ value of 0.01 in other cases. This results gain confirm the finding in the case of single subset: the introduction of considering for latent space distribution improves the system performance but the weight of distribution loss in latent space should remain fairly insignificant.

\section{Discussion}
\label{sec:discussion}

Although the proposed method uses reconstruction loss as the anomaly score, parameters related to latent vectors such as model likelihood can also be used as the anomaly score. We first propose an examination on the relationship of reconstruction loss and KL divergence between the latent vectors and the data distribution in the latent space. Pearson's correlation coefficient between the KL divergence and reconstruction loss are calculated. Suppose $l_\text{KL}^i$ represents the KL divergence between the the $i$th sample in the latent space and normal distribution, $l_r^i$ represents the reconstruction loss of the $i$th sample in terms of MSE, the Peason's correlation coefficient $\rho$ can be calculated as:

\begin{equation}
	\rho = \frac{\sum_i(l_\text{KL}^i-mean(l_\text{KL}))(l_r^i-mean(l_r))}{\sqrt{\sum_i(l_\text{KL}^i-mean(l_\text{KL}))^2}\sqrt{\sum_i(l_r^i-mean(l_r))^2}}.
\end{equation}

\begin{table}[htb]
	\centering
	\begin{tabular}{|c|c|c|c|c|c|c|}
		\hline
		Beta & 0 & 0.01 & 0.1 & 1 & 10 & 100  \\ 
		\hline
		$\rho_p$ & 0.44 & 0.41 & 0.32 & 0.79 & 0.79 & 0.79 \\
		\hline
	\end{tabular}
	\caption{The Pearson's correlation coefficients between the reconstruction loss and the KL divergence between the latent vector of sample and normal distribution. The results are calculated by the single subset system using subset `e'.}
	\label{tab:correlation}
\end{table}

Setting the same set of beta value, the correlation between KL divergence and reconstruction loss is shown in Table \ref{tab:correlation}. The correlation between reconstruction loss and the KL divergence between samples and normal distribution in the latent space become high with a large $\beta$ value. As a result, in the systems having a larger beta value, the reconstruction loss serving as the anomaly score should have a similar performance with the cases that parameters related to latent vectors are used as the anomaly score. Therefore the better performance of systems with smaller beta values then the systems with larger values is likely to suggest that the reconstruction based anomaly score may outperform the latent vector based anomaly score in this case of unsupervised PCG anomaly detection.

\section{Conclusion}
\label{sec:conclusion}

This paper proposes an unsupervised PCG analysis system that detects anomaly PCG signals by learning normal PCG signals only. The system is based on $\beta-$VAE system with the beta value of the best performed system is even smaller than 1 in most cases. Further investigations suggests that for this PCG analysis case, the anomaly score calculation has a small preference on reconstruction based parameters.

\bibliographystyle{IEEEbib}
\bibliography{Ref}

\end{document}